\documentclass[12pt]{article}

\usepackage{epsfig}
\def\beq{\begin{equation}}
\def\eeq{\end{equation}}

\def\tev{\, {\rm TeV}}
\def\gev{\, {\rm GeV}}

\newcommand{\gsim}{\lower.7ex\hbox{$\;\stackrel{\textstyle>}{\sim}\;$}}
\newcommand{\lsim}{\lower.7ex\hbox{$\;\stackrel{\textstyle<}{\sim}\;$}}

\begin{document}

%#!latex

%@twocolumnfalse\endcsname
%%
%%
\begin{titlepage}
\noindent
\begin{flushright}
%DRAFT 
%\today \\
MCTP-02-72\\
\end{flushright}
\vspace{1cm}

\begin{center}
  \begin{large}
    \begin{bf}
Revisiting Top-Bottom-Tau Yukawa Unification \\
in Supersymmetric Grand Unified Theories
    \end{bf}
  \end{large}
\end{center}
\vspace{0.2cm}
\begin{center}
Kazuhiro Tobe and James D. Wells\\
  \vspace{0.2cm}
  \begin{it}
	MCTP, Department of Physics,
	University of Michigan,
	Ann Arbor, MI 48109, \\
	Department of Physics, 
	University of California, 
	Davis, CA 95616
  \end{it}

\end{center}

%\maketitle
\begin{abstract}

Third family Yukawa unification, as suggested by minimal $SO(10)$ unification,
is revisited in light of recent experimental measurements and theoretical
progress.  We characterize unification in a semi-model-independent fashion,
and conclude that finite $b$ quark mass corrections from superpartners
must be nonzero, but much smaller than naively would be expected.
We show that a solution that does not require cancellations of
dangerously large $\tan\beta$ effects in observables implies that
scalar superpartner masses should be substantially
heavier than the $Z$ scale, and perhaps inaccessible to all currently
approved colliders.  On the other hand, gauginos must be
significantly lighter than the scalars.  
We demonstrate that a spectrum of 
anomaly-mediated gaugino masses and
heavy scalars works well as a theory compatible with third
family Yukawa unification and dark matter observations.

\end{abstract}

\vspace{1cm}

\begin{flushleft}
hep-ph/0301015 \\
January 2003
\end{flushleft}

\end{titlepage}
%\pacs{PACS numbers: }
%]

%%%%%%%%%%%%%%%%%%%%%%%%%%%%%%%%%%%%%%%%%%%%%%%%%%%%%%%%%%%%%%%%%%%%%%
\section{Introduction}

The unification of the gauge couplings in supersymmetric (SUSY) 
theories leads
one to entertain the possibility that the three gauge groups of
the Standard Model unify into a simple grand unified group.
Computations within this framework have shown that the low-energy
gauge couplings are not only compatible with supersymmetric unification,
but are predicted accurately to within expected slop of a sensible
high-scale theory.

Any unified theory of the leptons and quarks gives a symmetry
reason for their masses once the unified theory is broken
and parameters are renormalized to the low scale where we 
take measurements. The pressing issue then is to use what we know
to posit more complete forms of the high-scale
theory in order to make more predictions and identify more tests
for consistency.

We have mild clues to the form of the high-scale theory. First, the
unification of gauge couplings works provided the hypercharge of each
state is normalized as though it came from $SU(5)$ breaking.  This makes
any simple group which has an $SU(5)$ subgroup up for 
primary consideration, such
as $SU(5)$, $SO(10)$, $E_6$, etc.  Second, the existence of neutrino masses
gives preference to theories that naturally incorporate the left and right
neutrinos into simple representations.  For this reason, we wish
to focus on $SO(10)$, which is the smallest rank symmetry group
that pays full respect to these two clues.

The simplest models of $SO(10)$ have complete unification of the
third family Yukawa couplings at the high scale.  This requirement
implies the well-known criterion that $\tan\beta$ should be large
(about 50)~\cite{Banks:1987iu}.

When $\tan\beta$ is large, the renormalization group evolution of
the theory parameters (gauge couplings and Yukawa couplings)
can only be done accurately by two-loop
numerical computations.  The numerics are somewhat messy, and so the
hope was originally that the issue of Yukawa coupling unification
could be elucidated by a compensatingly easy form of supersymmetry
with few parameters.  For example, one would have like to have seen
how it all works for ``minimal supergravity'', where all the gauginos
have the same mass at the high scale and all scalars have the same
mass at the high scale.  Unfortunately, this simple model fails
to simultaneously allow Yukawa unification and electroweak symmetry
breaking~\cite{Murayama:1995fn}.

The lack of simple analytics to analyze Yukawa unification and the failure
of a simple model to exemplify it has created a little confusion (for us,
at least) on whether third family Yukawa unification is a reasonable
expectation.  The question has been studied 
by several groups in the past (for very recent papers, see
Refs.\cite{Blazek:2002ta,Baer:2001yy,Komine:2001rm}); however, we are
revisiting it because 
experimental input data on the strong coupling, the top mass, and the
bottom mass has improved dramatically in the past few years, there
are some discrepancies in the $b-\tau-t$ unification
literature~\cite{Blazek:1997cs,Baer:2000jj,Blazek:2002ta,Baer:2001yy},
and the theoretical palette of supersymmetric models has grown. We also
believe we have an effective way of showing in a semi-model-independent
fashion what the requirements, and challenges, are for Yukawa unification,
and we believe we have identified a general approach to the superpartner
spectrum that is worth considering as a solution to the Yukawa unification
problem. We illustrate the effectiveness of this approach within anomaly
mediated supersymmetry breaking.

%%%%%%%%%%%%%%%%%%%%%%%%%%%%%%%%%%
\section{Ultraviolet and infrared sensitivity of prediction}

When $\tan\beta$ is very high there is an extreme sensitivity of low-energy
parameters on the high-scale prediction of Yukawa unification.  This might
sound vaguely disturbing at first, but in fact it is an opportunity we
can exploit to narrow the low-scale options that work.  
In contrast,
extreme ultraviolet sensitivity would be disastrous to progress since 
it would not matter what shuffling of parameters we do at the low scale,
their effects would all be dwarfed by high-scale uncertainties.

We will demonstrate the infrared and ultraviolet sensitivity of the
predictions below.
We also hope to demonstrate a somewhat easier method to characterize
the success or failure of third family Yukawa coupling unification.
In the end, we hope we will have demonstrated two important conclusions
to the reader: third family Yukawa unification is compatible
with low-scale supersymmetry, but the low-scale finite 
corrections to the bottom quark must be smaller than naive expectations would
have them be.  This latter point translates into the requirements that
large low-scale SUSY threshold corrections must either cancel to give
a moderate or small-sized correction to the $b$-quark mass, or they
must be greatly suppressed.  When we discuss additional constraints on the
low-scale spectrum from experiment, the suppression explanation will
look more tenable in our view.

%%%%%%%%%%%%%%%%%%%%%%%%%%%%%%%%%%%%%%%%%%%%%%%%%%%%%%%%%%%%%%%%%%%%%
%\section{Parameterizing unification requirements}

In order to understand how important the low-energy SUSY corrections 
are for Yukawa unification, we first show approximate expression of the
GUT couplings, which depend on the low-energy SUSY corrections. The expressions
for the gauge couplings are
\begin{eqnarray}
g_1(M)&\simeq&0.73 \left( 1+3\delta_{g_1}-0.007 \delta_{g_2}
+0.02 \delta_{g_3} 
-0.02 \delta_{y_t}-0.005 \delta_{y_b}
\right.\nonumber \\
&&\left. 
-0.002 \delta_{y_\tau}
-0.007 \delta_{\tan\beta} +0.02 \log{\frac{M}{M_{G_0}}}
+\delta_{g_1}^{GUT}+O(\delta^2)\right),\\
%
%g_1(M)&\simeq&0.73 \left( 1-0.02 \delta_{y_t}-0.005 \delta_{y_b}
%-0.002 \delta_{y_\tau} \right.\nonumber \\
%&&\left.-0.007 \delta_{\tan\beta} +0.02 \log{\frac{M}{M_{G_0}}}
%+O(\delta^2)\right),\\
%
g_2(M) &\simeq& 0.73 \left( 1-0.003 \delta_{g_1}+ \delta_{g_2}
+0.03 \delta_{g_3}-0.02\delta_{y_t}-0.008\delta_{y_b} \right.
\nonumber \\
&&\left. -0.001\delta_{y_\tau}-0.01\delta_{\tan\beta}
+0.004\log{\frac{M}{M_{G_0}}} +\delta_{g_2}^{GUT}+O(\delta^2)\right),\\
%
%g_2(M) &\simeq& 0.73 \left( 1-0.02\delta_{y_t}-0.008\delta_{y_b} 
%-0.001\delta_{y_\tau}\right.\nonumber \\
%&&\left. -0.01\delta_{\tan\beta}
%+0.004\log{\frac{M}{M_{G_0}}} +O(\delta^2)\right),\\
%%
g_3(M) &\simeq & 0.72 \left(1-0.001\delta_{g_1}-0.002\delta_{g_2}
+0.4\delta_{g_3}-0.01 \delta_{y_t}-0.005\delta_{y_b} \right.
\nonumber \\
&&\left. -0.0002\delta_{y_\tau}-0.005\delta_{\tan\beta}
-0.01\log{\frac{M}{M_{G_0}}}+\delta_{g_3}^{GUT}+O(\delta^2) \right),
%
%g_3(M) &\simeq & 0.72 \left(1-0.01 \delta_{y_t}-0.005\delta_{y_b} 
%-0.0002\delta_{y_\tau}\right.
%\nonumber \\
%&&\left. -0.005\delta_{\tan\beta}
%-0.01\log{\frac{M}{M_{G_0}}}+O(\delta^2) \right),\\
%
\end{eqnarray}
where the scale $M$ should be close to the GUT scale $M_{G_0}\simeq 3\times
10^{16}$ GeV, defined to be
the location where $g_1(M_{G_0})=g_2(M_{G_0})$.
Here $\delta^{GUT}$ is a threshold correction at GUT scale, and
\begin{eqnarray}
\delta_{g_i} & \equiv & \frac{\bar{g}_i^{MSSM}-\bar{g}_i^{SM}}
                                 {\bar{g}_i^{SM}}, \\
\end{eqnarray}
are deviations from the SM gauge couplings at $m_Z$ due to
SUSY contributions.  The uncertainties in the gauge coupling measurements
at the $Z$ scale are also included in $\delta_{g_i}$ and must be taken
into account when considering the predictions for unification. This
is particularly true for the strong coupling constant $g_3$.

The parameterization of Yukawa threshold corrections is
\begin{eqnarray}
\delta_{y_i} & \equiv & \frac{\bar{y}_i^{MSSM}-\bar{y}_i^{SM}/\sin\beta_0}
                                 {\bar{y}_i^{SM}/\sin\beta_0}
\end{eqnarray}
which keeps track of 
the deviations from the naive two Higgs doublet Yukawa couplings
at $m_Z$ due to SUSY contributions ($\tan\beta =50$).  
Similar to the gauge coupling
parameters $\delta_{g_i}$, the Yukawa unification parameters
take into account also the uncertainty of the fermion mass measurements.
This is particularly important for $m_b$ and $m_t$, as their uncertainties
are comparable to SUSY threshold corrections.
Note that $\delta_{y_i}=\delta_i$ ($i=b,t,\tau$)
as defined by Eqns.~(\ref{DR-yukawa})-(\ref{yukawa_correction}) in the
Appendix.~\footnote{
Our SUSY threshold corrections to Yukawa couplings contain
 logarithmic corrections from the wave function renormalization of the
fermions as well as the finite corrections.}
We have also defined
\begin{eqnarray}
\delta_{\tan\beta} & \equiv & \frac{\tan\beta - \tan\beta_0}{\tan\beta_0}
\end{eqnarray}
where $\tan\beta_0=50$.

%-----------------------------------

We can do a similar exercise for Yukawa couplings at a high-scale $M$ near
$M_{G_0}$ to see qualitative features of Yukawa unification.  Many numerical
coefficients are highly sensitive to $\tan\beta$ and we have therefore
expanded our results around $\tan\beta_0=50$ since we know that is
near what $\tan\beta$ needs to be for third family unification to occur.
The Yukawa couplings are then given by
\begin{eqnarray}
y_t(M) &\simeq& 0.63 \left(1+0.9\delta_{g_1}+3\delta_{g_2}
-3\delta_{g_3} +7\delta_{y_t} +0.7 \delta_{y_b}\right.
\nonumber \\
&&\left.+0.02 \delta_{y_\tau}+0.7 \delta_{\tan\beta}
-0.01 \log{\frac{M}{M_{G_0}}} +\delta_t^{GUT}+O(\delta^2)\right),\\
%
%y_t(M) &\simeq& 0.63 \left(1+7\delta_{y_t} +0.7 \delta_{y_b}
%+0.02 \delta_{y_\tau}\right.
%\nonumber \\
%&&\left.+0.7 \delta_{\tan\beta}
%-0.01 \log{\frac{M}{M_{G_0}}} +O(\delta^2)\right),\\
%
y_b(M) &\simeq& 0.44 \left(
1+0.7\delta_{g_1}+2\delta_{g_2} -2\delta_{g_3}
+\delta_{y_t}+3\delta_{y_b}+0.2\delta_{y_\tau}
\right.\nonumber \\
&&\left.+3\delta_{\tan\beta}-0.02 \log{\frac{M}{M_{G_0}}} 
+\delta_b^{GUT}+O(\delta^2)\right),\\
%
%y_b(M) &\simeq& 0.44 \left(
%1+\delta_{y_t}+3\delta_{y_b}+0.2\delta_{y_\tau}
%\right.\nonumber \\
%&&\left.+3\delta_{\tan\beta}-0.02 \log{\frac{M}{M_{G_0}}} 
%+O(\delta^2)\right),\\
%
y_\tau(M) &\simeq& 0.52 \left(1+0.1\delta_{g_1}+\delta_{g_2}
-0.6\delta_{g_3}+0.2\delta_{y_t}+\delta_{y_b}
+2\delta_{y_\tau} \right. \nonumber \\
&&\left. +3\delta_{\tan\beta}-0.005\log{\frac{M}{M_{G_0}}} 
+\delta_\tau^{GUT}+O(\delta^2) \right),
\end{eqnarray}

%-------------------------------------------

These expression should not be used in quantitative analysis because
$O(\delta^2)$ corrections to Yukawa couplings are not negligible. But
they do demonstrate interesting qualitative features.
For gauge couplings, unification is not terribly sensitive
to low energy SUSY corrections (i.e., the coefficients of
$\delta_{y_i}$ and $\delta_{g_i}$ are small). 
On the other hand, Yukawa couplings at the GUT scale
are very sensitive to the low-energy SUSY corrections. An $O(1\%)$ correction
at low energies can generate close to a $O(10\%)$ correction at the GUT scale.
This extreme IR sensitivity is one source of the variance in conclusions in 
the 
literature~\cite{Blazek:1997cs,Baer:2000jj,Blazek:2002ta,Baer:2001yy}.  
For example, course-grained scatter plot methods, which are so useful in
other circumstances, lose some of their utility when IR sensitivity is so high.
Furthermore, analyses
that use only central values of measured fermion masses do not give a full
picture of what
range of supersymmetry parameter space enables third family Yukawa
unification, since small deviations in low-scale parameters can mean
so much to the high-scale theory viability.

% --------- Start altering here again ---------------------------
\section{Semi-model-independent analysis of corrections}

We can make progress in understanding unification of the Yukawa couplings
prior to positing a specific set of supersymmetric masses, and even
prior to agreeing on a supersymmetry breaking framework.  
The technique we utilize is to consider the SUSY corrections 
$\delta_{t,b,\tau}$ as free parameters. We vary $\delta_{t,b,\tau}$
and $\tan\beta$ at $m_Z$, and search for $b-\tau-t$ Yukawa coupling
unification at the GUT scale. For this analysis we define the
GUT scale to be the scale where $g_1$ and $g_2$ unify. 
For the SUSY corrections to gauge couplings $\delta_{g_i}$, we use
the formulas Eqns.~(\ref{MSSM_gauge_cc}) and~(\ref{MSSM_gauge_corr}) 
in the Appendix,
assuming that all SUSY particle masses are $1$ TeV. Later, we will also
discuss the dependence of $\delta_{g_3}$.

\begin{figure}[thb]
\centering
\includegraphics*[width=7.9cm]{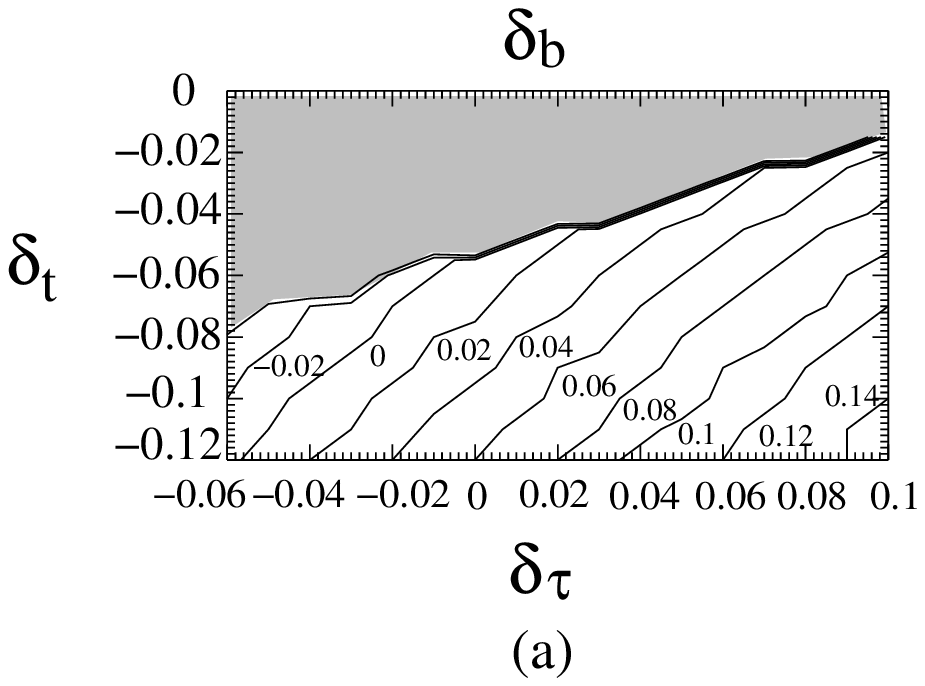}
\includegraphics*[width=7.9cm]{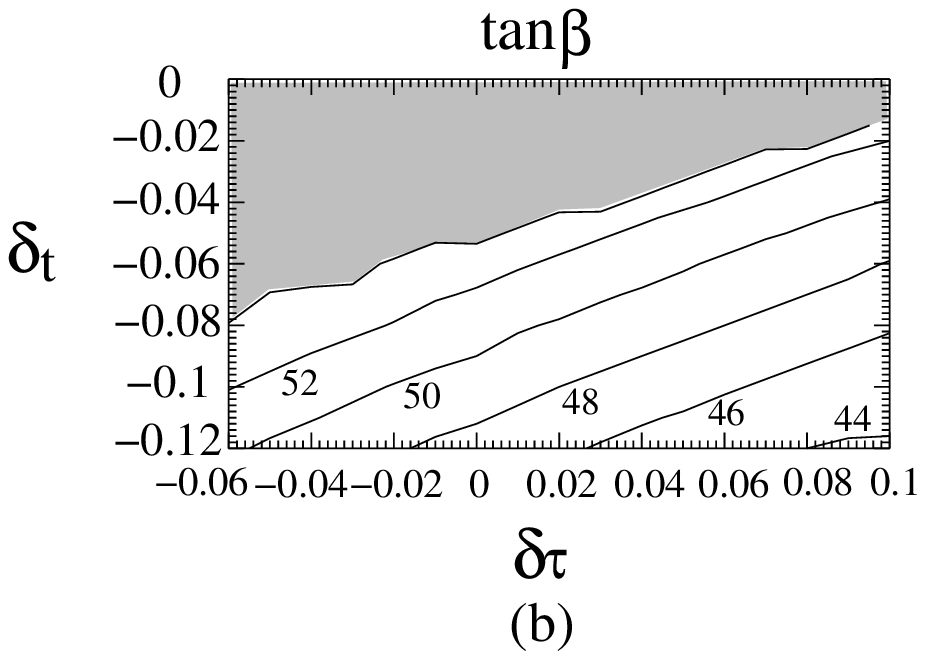}
\caption{(a) SUSY threshold corrections $\delta_b$ to the bottom Yukawa
coupling at $m_Z$, required for the Yukawa coupling unification,
as a function of the SUSY threshold corrections to the top
Yukawa coupling ($\delta_t$) and tau Yukawa coupling ($\delta_\tau$).
(b) Contours of $\tan\beta$ needed for Yukawa unification in the
parameter space of (a). The shaded region is where the needed value
of $\tan\beta$ gets so high that the $b$ Yukawa
coupling goes non-perturbative below the GUT scale.}
\label{yukawa_unif_condition}
\end{figure}

Requiring exact Yukawa unification, we get the constraints among
$\delta_{t,b,\tau}$ shown in Fig.~\ref{yukawa_unif_condition}a.
The contours are of $\delta_b$ values such that
given a $\delta_\tau$ value ($x$ axis) and a $\delta_t$ value ($y$ axis)
we get equality of the Yukawa couplings at the GUT scale.
Fig.~\ref{yukawa_unif_condition}b reveals the $\tan\beta$ value needed
to get unification for the given set of $\delta_f$ values.  As expected,
the allowed values of $\tan\beta$ are quite high and are consistent
with previous analyses.

%--------------------------------------------------------------
%--------------------------------------------------------------

The size of corrections for $\delta_t$ and $\delta_\tau$ in typical
weak scale supersymmetric theories with $\tan\beta\sim 50$ are roughly
\begin{eqnarray}
|\delta_t| &\simeq&
 \left|\frac{g_3^2}{6\pi^2}\log\left(\frac{m_Z}{M_{\rm SUSY}}\right)\right|
\lsim 10\%,\nonumber \\ 
\delta_\tau &\sim& \frac{g_2^2}{32 \pi^2}\frac{M_2\mu\tan\beta}{M^2_{\rm
SUSY}}\lsim {\rm few}\%. 
\end{eqnarray}
Therefore, 
one important point to notice in these graphs is the relatively small 
corrections allowed for $\delta_b$ compared to the expectation of
$|\delta_b|\sim (g_3^2/12 \pi^2)\tan\beta\sim
50\%$  for high-$\tan\beta$
supersymmetric theories with masses less than
about a TeV and gauginos and sfermions roughly equal in mass. 

\begin{figure}[thb]
\centering
\includegraphics*[width=7.9cm]{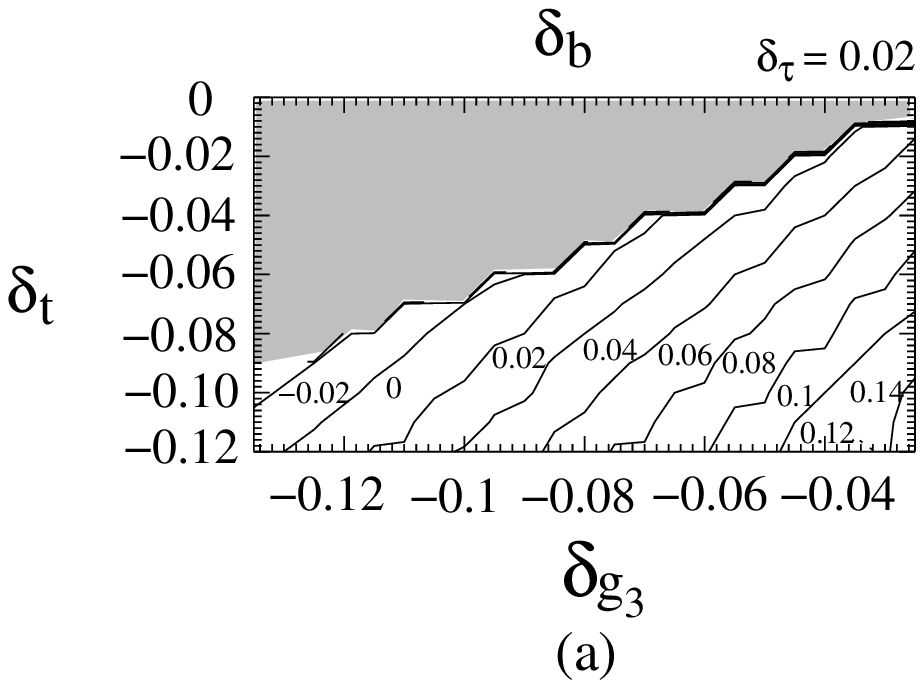}
\includegraphics*[width=7.9cm]{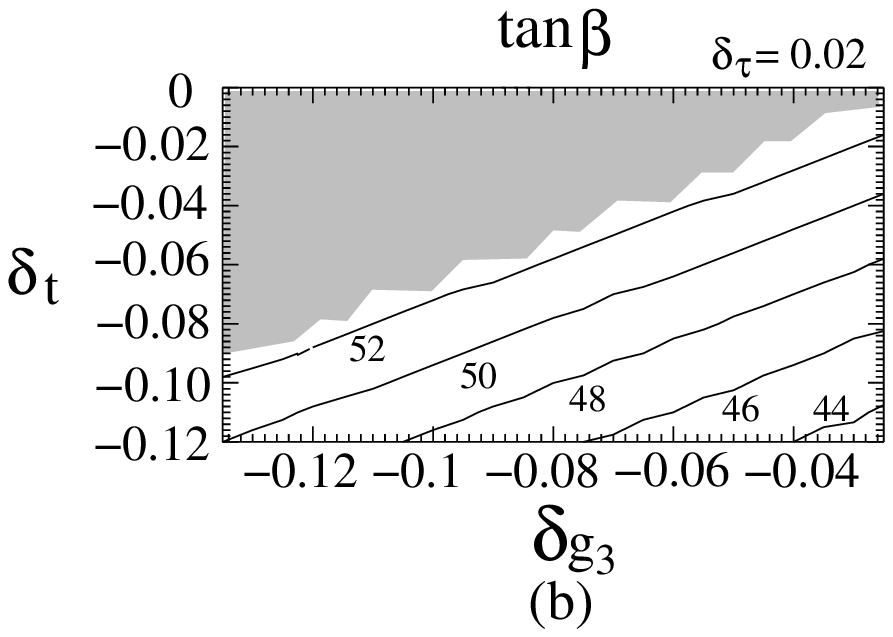}
\caption{(a) SUSY threshold corrections $\delta_b$ at $m_Z$ required for
Yukawa coupling unification, as a function of the top Yukawa SUSY
threshold corrections ($\delta_t$) and SUSY threshold corrections to 
the strong gauge coupling ($\delta_{g_3}$). Here we fix $\delta_\tau=0.02$.
(b) Contours of $\tan\beta$ to achieve Yukawa coupling unification
in the parameter space of (a). The shaded region is where the needed value
of $\tan\beta$ gets so high that the $b$ Yukawa
coupling goes non-perturbative below the GUT scale.}
\label{yukawa_unif_condition2}
\end{figure}

The uncertainties and corrections in the strong coupling constant
($g_3$) also
play a role in the analysis, but do not change the qualitative
picture.  We expect the combination of weak-scale
SUSY threshold corrections and the measurement
uncertainty to be below about $|\delta_{g_3}|\lsim 10\%$.  From
Fig.~\ref{yukawa_unif_condition2} we see that the values of the
$b$ quark corrections are still required to be small,
$\delta_b\lsim 10\%$, and $\tan\beta$ is still near 50.

%--------------------------------------------------------------
%-------------------------------------------------------------

Exact unification of the renormalized weak-scale Yukawa couplings
at the high scale is not expected. The renormalized weak-scale
Yukawa couplings will flow up to the high scale and be slightly mismatched,
but the high-scale threshold corrections will shift them such that they
unify.  We therefore only need to require that the renormalized couplings
flow within a reasonable neighborhood of each other, where reasonable
neighborhood is defined to be size of the mismatch expected from the
high-scale threshold corrections.  
In our above analysis, we have not included neutrino Yukawa coupling effects.
If the tau neutrino Yukawa coupling is also unified with others at GUT
scale, the right-handed neutrino mass scale $(M_R)$ should be around
$10^{13}-10^{15}$ GeV to explain atmospheric neutrino mass scale
(assuming hierarchical neutrino masses). Therefore the right-handed
neutrino scale is close to GUT scale. Here we take the neutrino
Yukawa running effect as a GUT threshold effect. From one loop 
$\beta$-functions, the corrections are estimated to be at most a few \%
in the positive direction:
\begin{eqnarray}
\delta_t^{GUT} &\simeq& \delta_\tau^{GUT} \simeq 
\frac{y_\nu^2}{16\pi^2}\log\frac{M_G}{M_R},\\
\delta_b^{GUT} &\simeq&0.
\end{eqnarray}

\begin{figure}
\centering
\includegraphics*[width=10cm]{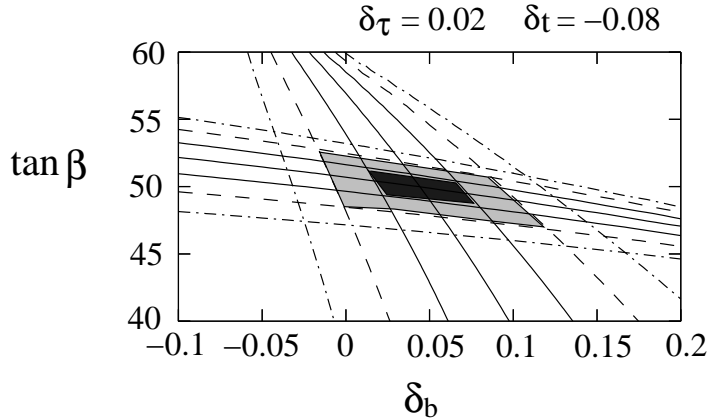}
\caption{Contours of $[y_b(M_G)-y_\tau(M_G)]/y_b(M_G)$ (vertical lines) and
$[y_t(M_G)-y_\tau(M_G)]/y_t(M_G)$ (horizontal lines) as a function of
$\delta_b$ and $\tan\beta$. We fix $\delta_\tau=0.02$
and $\delta_t=-0.08$. The dark-shaded region is the
allowed region for Yukawa coupling unification, given $\pm 5\%$
threshold corrections at the GUT scale. The bigger light-shaded region is
for $\pm 10\%$ threshold corrections.} 
\label{yukawa_unif2}
\end{figure}

It is hard to anticipate all high-scale
threshold corrections, but it has been argued that Yukawa corrections
at the high-scale are small, not more than a 
few percent~\cite{Blazek:2002ta}.  To illustrate
the effect this has on the low-scale theory predictions for 
$\tan\beta$ and the bottom quark mass corrections, we have chosen
typical values of $\delta_t$ and $\delta_\tau$ weak-scale threshold
corrections and plotted in  Fig.~\ref{yukawa_unif2} contours of $\tan\beta$
and $\delta_b$ for various sizes of high-scale threshold corrections.

The center vertical solid line in the figure corresponds to no high-scale
threshold corrections to $[y_b(M_G)-y_\tau(M_G)]/y_b(M_G)$.
The center horizontal solid line in the figure corresponds to no high-scale
threshold corrections to $[y_t(M_G)-y_\tau(M_G)]/y_t(M_G)$.
The various other lines are spaced at 5\% threshold correction increments
as one moves away from the center line.  The solid parallelogram
in the middle of the figure is the allowed region for $\tan\beta$ and
$\delta_b$ given 5\% threshold corrections at the high-scale.  The 
bigger shaded region is for 10\% threshold corrections.  These reasonable,
and perhaps even large, high-energy threshold corrections still do not
change the basic conclusion that we would like to make here:
$\delta_b$ corrections are required to be small.

%------------------------------------------------------------------
%------------------------------------------------------------------

Remember,
in the construction of these graphs no assumptions have been made on the
superpartner spectrum, so it will be up to us now to find out what 
spectrum can accommodate these corrections.  More to the point: how
can we get small $\delta_b\sim 5\%$ with such a large 
$\tan\beta\sim 50$?

%%%%%%%%%%%%%%%%%%%%%%%%%%%%%%%%%%%%%%%%%%%%%%%%%%%%%%%%%%%%%%%%%
\section{Superpartner spectrum requirements}

We have made statements in the previous section that when $\tan\beta\sim 50$
and superpartner masses are below about $1$ TeV
we expect $b$ quark mass corrections to be much
higher than the $\delta_b\sim 5\%$
required for third family unification.  A quick way to see this is in
the contributions to the finite corrections which scales as
\beq
\delta^{\rm finite}_b \simeq -\frac{g_3^2}{12\pi^2}
 \frac{\mu M_{\tilde g}\tan\beta}{m_{\tilde b}^2} 
+ \frac{y_t^2}{32\pi^2}\frac{\mu A_t \tan\beta}{m_{\tilde t}^2} +\ldots
\eeq
If all superpartner masses are about the same, we see that each term
contributes about
$\delta_b\sim 50\%$~
\cite{Hempfling:1993kv,Hall:1993gn,Carena:1994bv,Rattazzi:1995gk,
Pierce:1996zz}   
with the sign depending on the relative sign
of $\mu M_{\tilde g}$ and $\mu A_t$.

To reduce the size of this correction to the $\delta_b\sim 5\%$ level
one needs to either cancel large individual terms in the formulas for
the corrections, or suppress the corrections.
In our view, the appeal of the first option is diminished, but not ruled out, 
when we contemplate the strong correlation 
between large SUSY contributions to $\delta_b$ and large
SUSY contributions to $b\to s\gamma$.  

We know from experiment
that the decay rate $b\to s\gamma$ is consistent with the Standard
Model~\cite{Kagan:1998ym,Ali:2002jg}, 
and weak-scale supersymmetry (masses less than about a TeV)
with large $\tan\beta$ supersymmetry generally is
wildly in conflict with it. There are two ways out of this problem: 
the large individual SUSY corrections, including $t-H^\pm$,
$\tilde t-\tilde \chi^\pm$ and $\tilde b-\tilde g$ loop corrections,
could conspire to cancel each other for a small total effect, or these
same loop corrections could be so large that they are almost exactly
twice the size of the SM amplitude, but opposite in sign so that 
the overall observable is not altered.  This possibility is still
allowed by the data, although future measurements of the 
FB asymmetry of $b\to sl^+l^-$ would distinguish the sign of the 
amplitude~\cite{Ali:2002jg}.

The terms in the $b\to s\gamma$ computation are very similar
to the terms in the $\delta_b$ computation.  They both involve loops
of charginos and stops, and loops of gluinos and sbottoms (if supersymmetric
CKM matrix is different than CKM matrix), but the coefficients and
couplings are 
different.  A cancellation of supersymmetric effects in one does not
by any means imply a cancellation in the other. Therefore,
it is perhaps reasonable for a low-scale supersymmetric
model with $\tan\beta\sim 50$
to conspire to give small corrections to either $b\to s\gamma$ or 
to $\delta_b$, but we view it as unlikely that cancellations occur
for both observables.

We are next led to the explanation that $\delta_b$ is small because
all terms contributing to it are suppressed.  It has been pointed out
that an approximate $R$ symmetry or PQ symmetry can systematically
suppress both $\delta_b$ and $b\to s\gamma$~\cite{Hall:1993gn}.      
Both computations go
to zero as supersymmetric parameters carrying $R$-symmetry charge
(gaugino masses and $A$ terms) go to zero, and both computations
go to zero if terms carrying PQ charge (the $\mu$ term) go to zero.

The constraints from $b\to s\gamma$ are very severe at $\tan\beta\sim 50$.
If all superpartner masses are less than 1 TeV it is challenging to
envision a superpartner spectrum capable of 
suppressing individual contributions enough to not require a
rather large beneficial cancellation (conspiracy of amplitude signs)
to occur in the finite $b$-quark to allow Yukawa unification.  And in 
addition, we simultaneously need equivalently large contributions
to the $b\to s\gamma$ amplitude to cancel or add with just the right
strengths to recover the magnitude of the SM amplitude but with opposite sign.

The problem of making relatively light supersymmetry compatible with
Yukawa coupling unification was taken up 
in the excellent study by Bla\v{z}ek, Derm\' \i \v sek and Raby
(BDR)~\cite{Blazek:2002ta}.\footnote{Similar findings for 
Yukawa unification were found in Ref.~\cite{Baer:2001yy} using a 
D-term model approach and assuming larger possible GUT scale
Yukawa corrections.}
The BDR approach is within a model of unified scalar
mass $m_{16}$ 
at the high scale for all states in the 16 representation (squarks,
sleptons, sneutrinos), split Higgs masses within the 10 representation,
and unified gaugino masses $m_{1/2}$ at the GUT scale.  They find 
that in order for third family Yukawa unification to occur the SUSY
parameters must have the following properties:
\begin{itemize}

\item[1.] $\tan\beta \sim 50$ for the third generation
Yukawa couplings to approach unification.

\item[2.] $m_{1/2}\sim \mu \ll m_{16}$  to suppress, but not to zero, 
the gluino contributions to
$\delta_b$ and $b\to s \gamma$.

\item[3.] Large $A_0$ such that the weak-scale $A_t$ is larger than
$M_{\tilde{g}}$ and the positive 
chargino-stop contributions to $\delta_b$ cancels, and slightly overcomes,
the large negative contributions due to gluino-sbottoms finite and logarithmic
corrections. (Note, we are expressing this criteria in our sign convention
for $b$-quark corrections which is opposite to BDR sign convention:
$\delta_b \propto -\Delta m_b^{\rm BDR}$).  

\item[4.] $\mu > 0$ so that the large chargino-stop 
corrections to $b\to s\gamma$ can be opposite in sign to SM (and charged
Higgs) contributions. This is necessary to be consistent with the
large choice of $A_0$ term above, which when combined with the right sign
of $\mu$ gives the
chargino-stop loops a sufficiently large canceling contribution to
change the sign of the $b\to s\gamma$ amplitude. This enables large 
$\tan\beta$ supersymmetry 
to be consistent with the $B(b\to s\gamma)$ measurements
despite the SUSY 
contributions being much larger than the SM contributions.

\end{itemize}
The above criteria for third generation Yukawa unification may seem
tortured to the uninitiated, but we can verify after much technical
work in hopes of finding 
something ``better'' that the BDR solution is the right approach
if supersymmetry is relatively light.  
It is remarkable how robustly unique of a general solution approach
it turns out to be, and we concur with BDR on its interesting phenomenological
predictions, such as $m_h\simeq 114\pm 6$ GeV, $m_{\tilde t_1}\ll
m_{\tilde b_1}$, etc.

There is one disquieting feature of the BDR solution that
one might call ``finetuned cloaking of large $\tan\beta$ effects.''
It is unsettling to expect
that there is a nice conspiracy of large effects coming out just
right (and more or less SM-like) for $b$-quark finite corrections
and $b\to s\gamma$ corrections.  Large corrections would not be a cause
for any concern whatsoever.  However, large corrections that cancel to 
yield small $b$-quark corrections and add just the right amount
to suppress or flip the sign of
the $b\to s\gamma$ amplitude such that we would see no deviation from
the SM may be too much to ask of a theory.   

There are other, perhaps just as severe, challenges to the BDR solution.
There may be another conspiracy involved in getting $g-2$ to work out properly.
However, as BDR have pointed out, it is generally less severe than 
$b\to s\gamma$ considerations, and the required 
choice of $\mu >0$ at least goes
in the preferred direction for $g-2$.  A more difficult challenge would be
to get acceptable dark matter solutions. When scalar masses are much larger
than gaugino masses, the relic abundance of the stable lightest neutralino
is typically much too large to be acceptable.  Additional tunings of the 
spectrum and parameters would be required to make this work out properly.
For example, just the right mixture of higgsino and bino components to the
lightest neutralino would guarantee any amount of dark matter one wishes,
but this requires a somewhat sensitive tuning between $m_{1/2}$ and $\mu$.

Annihilations of the lightest neutralino through a heavy Higgs
pole is another viable option to suppress relic abundance in high
$\tan\beta$ theories of this sort~\cite{Baer:2001yy}, but 
it requires the heavy Higgs states to be relatively
light in contradistinction to other scalars in the theory. Nevertheless,
the large $\tan\beta$ suppression of the pseudoscalar mass with respect
to lagrangian parameters is encouraging for this scenario.

%%%%%%%%%%%%%% New Solution ****************************************
\section{The partially decoupled solution}

We would like to discuss another solution to third generation
Yukawa unification that involves no finetuned cloaking of the
large $\tan\beta$ effects.  As the overall scale of supersymmetry breaking
increases, the supersymmetric effects in $b\to s\gamma$ (and $g-2$)
decouple to zero.  However, the finite and logarithmic threshold corrections
to the Yukawa couplings do not decouple.

Logarithmic corrections to $\delta_b$ and $\delta_t$ are negative and
can be over 10\% for squark masses above several TeV.  The logarithmic
corrections to $\delta_\tau$ are negligible (less than 1\%).  Large
negative corrections to $\delta_b$ go in the wrong direction for Yukawa
unification, and if they stood alone would invalidate the hypothesis
of third generation Yukawa unification consistent with reasonable
high-scale threshold corrections. 

Once we realize that there are negative and large irrepressible
logarithmic corrections to $\delta_b$ and $\delta_t$
it becomes clear that
moderately valued
finite corrections (about 10\% to 20\%)
to $\delta_b$ from gluino-squark loops must be present.  
This is a very reasonable request of the theory since the $\delta_b$
corrections depend only on ratios of masses and not on the overall scale
of supersymmetry breaking: $\delta_b\propto -\mu M_{\tilde g}/m_{\tilde b}^2$.
This non-decoupling property of the finite $b$ quark mass corrections
is what allows us to suppress large $\tan\beta$ effects in potentially
dangerous observables and yet get large enough finite $b$-quark mass
corrections to obtain viable third family Yukawa unification.

The solution described above is transparently natural in every way except
electroweak symmetry breaking might be finetuned because we need large
scalar masses.  For example, stop masses above a few TeV are needed for
this solution to be realized in a way that naturally satisfies
large $\tan\beta$ sensitive observables, such as $b\to s\gamma$.

We have identified anomaly mediated gaugino 
masses and $A$ terms~\cite{Randall:1998uk,Giudice:1998xp} as
a theoretical scenario which may naturally realize
this approach to third generation Yukawa unification. We utilize several
features of this idea to our benefit.  If no singlets are around to
transmit supersymmetry breaking, the gaugino masses and $A$ terms
(dimension 3 SUSY breaking terms) may primarily arise
via anomaly mediation and are approximately one-loop suppressed compared
to the gravitino mass, and hence the $A$ terms are naturally the size of 
the gaugino masses. 

We also remind the readers that although the gaugino
mass relations in anomaly mediation do not look like they have any
connection to a GUT theory, they are in fact perfectly consistent
with a GUT unified gaugino mass.  The odd splitting of gaugino masses
below the GUT scale comes from large gauge-mediated threshold
corrections from integrating out heavy GUT states. The ``magic'' of
anomaly mediation is that these corrections are precisely what is needed
to recover the scale-invariant expressions for the gaugino masses
at low energy.

On the other hand, the scalars can naturally get
masses of the order the gravitino mass because $F^\dagger F/M_{\rm Pl}^2$
scalar mass-squareds do not require singlets. Scalar masses may be 
suppressed somewhat with respect to gravitino mass if the K\"ahler potential
approaches the no-scale form~\cite{Randall:1998uk,Giudice:1998xp}.  
We therefore make the general observation
that anomaly mediation is most naturally manifest by gauginos much lighter
than scalars, although the precise ratio between them depends on a priori
incalculable aspects of the theory (i.e., the K\"ahler potential).

To illustrate how Yukawa unification works in this scenario we normalize
the anomaly mediated gaugino masses to $M_2=150$ GeV.  This is well above
the experimental limits. (Recall that $\chi^\pm-\chi^0_1$ near degeneracy
makes experimental
detection extremely challenging, and normal chargino limits do 
not apply~\cite{Gunion:2001fu}.) This implies that $M_1\simeq 500\gev$ and 
$M_{\tilde g}\simeq -1300\gev$. It also implies that the gravitino
mass should be about $60\tev$.

\begin{figure}
\centering
\includegraphics*[width=10cm]{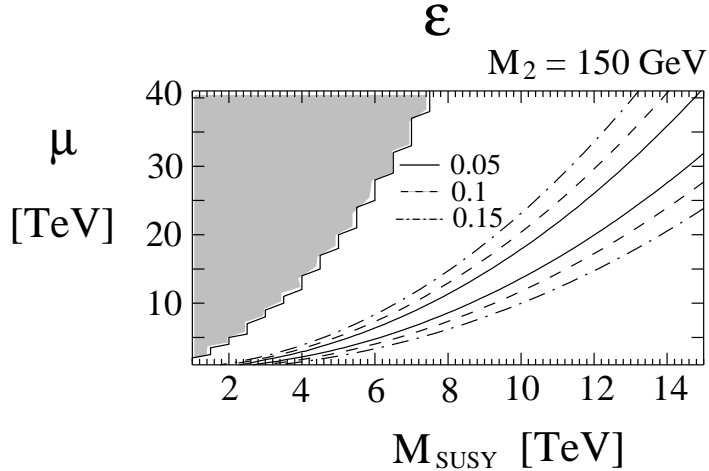}
\caption{Contours of $\epsilon$ defined by Eq.~(\ref{epsilon}).
$\epsilon=0.05~ (0.1, 0.15)$ corresponds to about $5 \%~ (10\%, 15\%)$
GUT threshold correction needed to achieve Yukawa coupling
unification. GUT-scale Yukawa corrections are expected to be 
less than about $1\%$. $M_{\rm SUSY}$ is the low-energy mass for all
scalar superpartners. The gaugino and A-term masses are equal to 
their anomaly-mediated values normalized to $M_2=150$ GeV.}
\label{M2_mu}
\end{figure}

In the next step we compute 
\beq
\epsilon =\sqrt{\left( \frac{y_b-y_\tau}{y_b}\right)^2
+\left( \frac{y_t-y_\tau}{y_t}\right)^2 
+\left( \frac{y_t-y_b}{y_t}\right)^2}~~~{\rm (computed~at~GUT~scale)}
\label{epsilon}
\eeq
as a function of the soft SUSY breaking sfermion masses
(which we assume all to be degenerate $M_{\rm{SUSY}}$)
and $\mu$.  Reasonable unification requires $\epsilon < 0.05$.
In Fig.~\ref{M2_mu} we plot contours of $\epsilon$ in the 
$M_{\rm SUSY}$--$\mu$ plane.  We note that if we rescale $m_{\tilde
b}(\simeq M_{\rm SUSY})$ by
a factor of $a$ then $\mu$ must be rescale by a factor of $a^2$ in order
to keep the finite $\delta_b\propto -\mu M_{\tilde g}/m_{\tilde b}^2$
constant for a given value of $M_{\tilde g}$.  Therefore, since we expect
$\mu$ to not be much heavier than $m_{\tilde b}$ we could gather 
from Fig.~\ref{M2_mu} that the squark and $\mu$-term masses should be
less than about $10\tev$ in this case.  Combining that with our requirement
that large $\tan\beta$ effects mostly decouple in $b\to s\gamma$ means
we expect
\beq
|M_{\tilde g}|\simeq 1300\gev\longrightarrow 
m_{\tilde b} \gsim {\rm few}\tev.
\eeq
In our case, the $A$ term value is quite small since it scales with
the gaugino masses in anomaly mediation. Furthermore, we have assumed
minimal flavor violation.  Relaxing that requirement would tighten
up the $b\to s\gamma$ very significantly, and push the required
$b$-squark masses beyond $5\tev$ or more.

%------------------------------------------------------------------
%------------------------------------------------------------------

In general the expected ratio of $m_{\tilde b}/M_{\tilde g}$ needs to
be between about 5 and 15 for Yukawa unification.  For scalar masses right
at the gravitino mass, this ratio would be about 45, so we need the scalar
mass to be at least a
factor of 4 suppressed compared to the gravitino mass.  Again, this presumably
would come from a K\"ahler potential that suppresses the mass, although
not nearly so absolutely, and perhaps not so unnaturally, 
as the no-scale K\"ahler potential.

\begin{figure}
\centering
\includegraphics*[width=7.9cm]{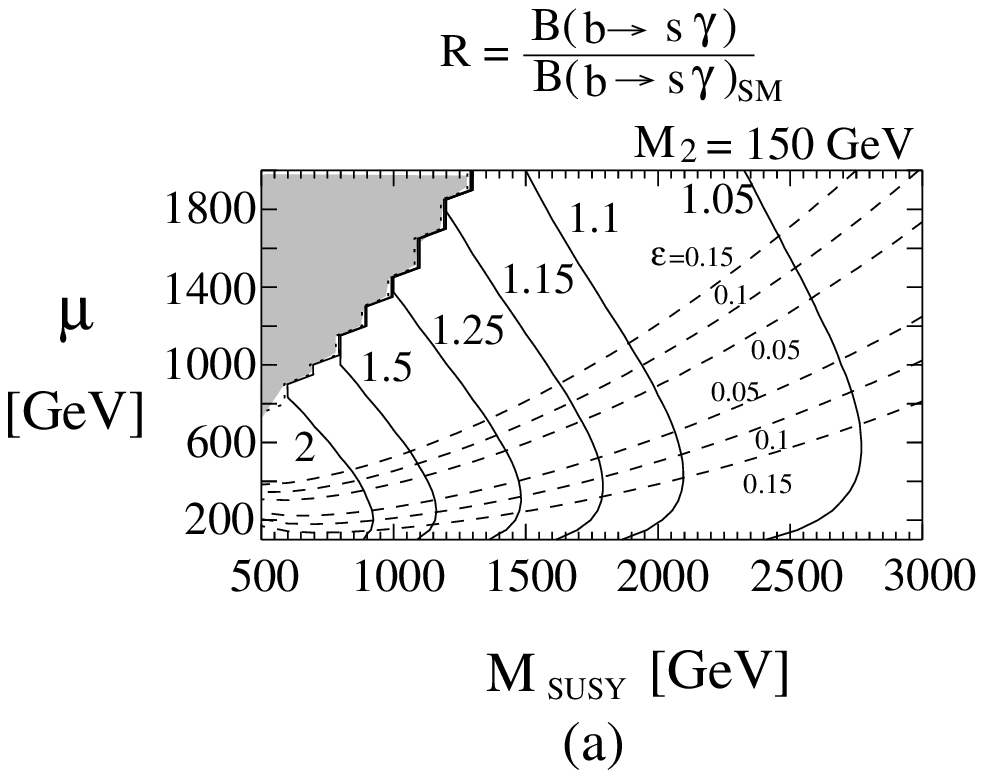}
\includegraphics*[width=7.9cm]{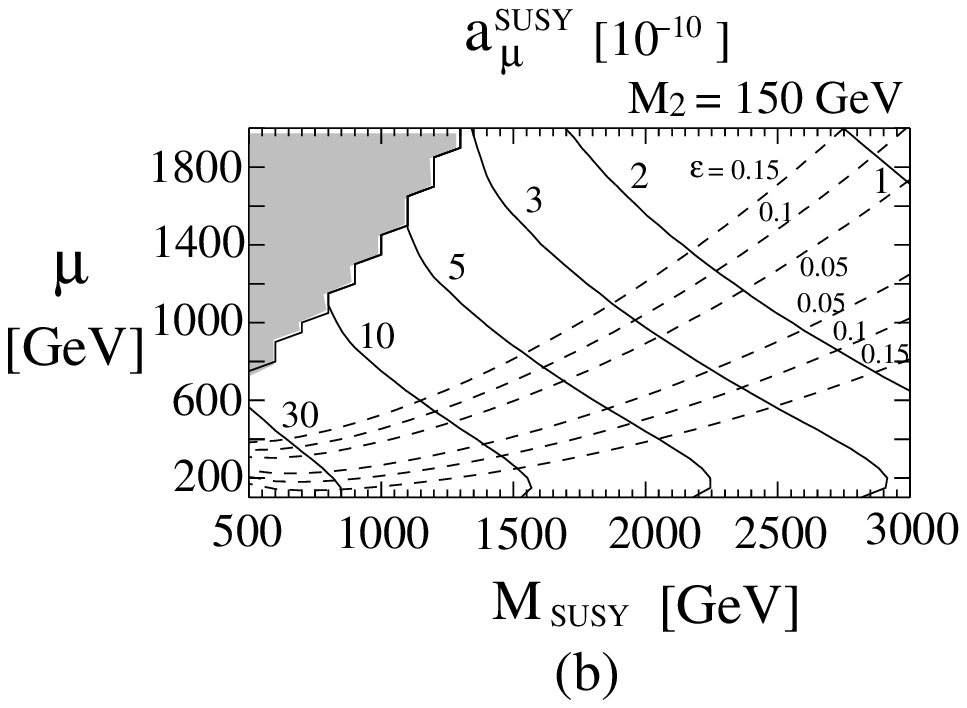}
\caption{(a) Contours of $R=B(b\rightarrow s\gamma)/B(b\rightarrow
s\gamma)_{\rm SM}$. 
(b) Contours of the SUSY contribution to muon $g-2$ ($a_\mu^{\rm SUSY}$).
In Fig.~(a) and (b), we also plot contours of
$\epsilon~(=0.05,0.1,0.15)$ from Fig.~{\ref{M2_mu}}.
$M_{\rm SUSY}$ is the low-energy mass for all
scalar superpartners. The gaugino and A-term masses are equal to 
their anomaly-mediated values normalized to $M_2=150$ GeV.
}
\label{bsg_M2}
\end{figure}

As stated earlier, 
we have computed the value of $B(b\to s\gamma)$ in this framework
by assuming the minimal flavor violation, and we have applied
the above supersymmetry spectrum to the 
formula of Ref.~\cite{Kagan:1998ym}.
We demonstrate in Fig.~\ref{bsg_M2} that indeed the
large $\tan\beta$ effects are easily decoupled in the interesting
range of scalar masses above a few TeV.  The figure plots contours
of constant
\beq
R=\frac{B(b\to s\gamma)}{B(b\to s\gamma)_{\rm SM}}.
\eeq
The data requires that this ratio be within 
$0.37 < R<1.25$~\cite{Kagan:1998ym}.
Our parameter space easily satisfies this bound. 

Furthermore, we have
checked that $g-2$ is easily consistent with the parameter space.
A conservative view of the $g-2$ experimental uncertainties and
theoretical uncertainties implies that $a_\mu^{\rm susy}/10^{-10}$
should be between about -37 and 90~\cite{Martin:2002eu}.  The predictions
for $g-2$ given in Fig.~\ref{bsg_M2}b are certainly well within that
range.   A less conservative interpretation of the experiment and
theory implies~\cite{Davier:2002dy} that 
\beq
a_\mu^{\rm susy}/10^{-10} = 34\pm 11~~~
(e^+e^-~{\rm based~analysis}).
\eeq
Therefore, there is a slight preference for positive values of
$g-2$ which our approach can accommodate.

Given the above analysis on $b\to s\gamma$ and $g-2$ of the muon, 
we feel 
comfortable concluding that Yukawa unification is consistent
with our solution to third generation Yukawa unification.

%------------------------------------------------------------------
%------------------------------------------------------------------

The drawback of our approach is that it increases
the apparent finetuning of radiative electroweak symmetry breaking since
all the scalars are at least several TeV.
We have made a trade compared to the BDR solution: 
we have rid ourselves of the several finetuned cloakings
of large $\tan\beta$ effects and traded it in for a larger finetuning
for EWSB. 

The anomaly-mediated case we presented here is also happily consistent
with the lightest supersymmetric particle being the dark matter.
In this case, the lightest supersymmetric particle
is a neutral Wino particle.  The relic abundance
calculated from thermal equilibrium evolution and freezeout of
Winos yields a much too low relic abundance to be of cosmological
interest.  This is of course less problematic than the overclosure
problem which faces a typical spectrum of the BDR solution.  Nevertheless,
it would have been disheartening if the perfectly good prospect of LSP dark
matter is no longer even viable. Fortunately this is not the case. Non-thermal
production of  the lightest neutralino 
via late-time gravitino~\cite{Gherghetta:1999sw} 
or moduli~\cite{Moroi:1999zb}
decays works surprisingly well and is easily compatible with the right amount
of cold dark matter needed to explain astrophysical observations.

%%%%%%%%%%%%%%%%%%%%%%%%%%%%%%%%%%%%%%%%%%%%%%%%%%%%%%%%%%%%%%%%%%
\section{Conclusion}

Supersymmetric grand unified theories are extremely attractive largely
because gauge coupling unification works so well.  The theory of
low-energy supersymmetry 
is both compatible with gauge coupling unification and it is a predictive
theory.  Compatibility just means it is possible, and predictivity means
it works surprisingly well -- e.g., the low-scale value of $g_3$
is predicted to within a few percent even after taking into account
the range of all reasonable high and low-scale threshold effects. 

Similarly, low-scale supersymmetry 
is certainly compatible with third generation Yukawa unification, 
and, to a lesser extent, it 
is predictive.  In this case, given that we now know all
the SM fermion masses rather well, we can predict
$\tan\beta$ and $\delta_b$ to interesting accuracies.  These can
then be traded in for knowledge/constraints on the
superpartner spectrum.

Our analysis argues that a viable solution to Yukawa unification 
which has no finetuned cancellation of large $\tan\beta$ effects
in observables requires very heavy scalar fields.
We would say loosely that  scalar
masses are above several TeV if this idea is what nature realizes.  We also
require that either the $\mu$ term or 
the $R$ charged masses (gaugino masses and $A$ terms)
must be light. A promising theory in this direction is anomaly mediated
supersymmetry breaking, where the gaugino masses are predominantly anomaly
mediated, and the scalar masses are somewhat suppressed compared to the
gravitino mass.  The compatibility with gauge coupling unification is
assured, compatibility with all dangerous
chirality-flipping observables is assured, and the LSP is retained as
a viable dark matter candidate.

Finally, we have not found a superpartner spectrum that all would agree
has no problems with finetuning.  In the BDR approach, which we have
verified as a viable solution, there is perhaps a finetuned cancellation
of large $\tan\beta$ effects in observables.  In our approach, there is
perhaps a finetuned potential for electroweak symmetry breaking.  
We are painted in these two corners primarily because of recent experimental
constraints (LEP, Tevatron, rare $B$ decays, precision electroweak, etc.). 
It would not be a totally indefensible position to say that experiment
in the last decade has diminished the viability of third generation
Yukawa unification.  Our analysis could even be subpoened to support
this view.  Nevertheless, we believe that the extreme infrared sensitivity
of Yukawa unification is a great opportunity to test the theory, and it
should be of no surprise that such an IR sensitive concept would get
severely squeezed shortly before it is redeemed or snuffed
out by direct superpartner measurements.

%%%%%%%%%%%%%%%%%%%%%%%%%%%
\section*{Acknowledgements}
K.T. thanks R.~Derm\' \i \v sek for useful discussions. 
K.T. and J.D.W. were supported in part by the U.S. Department of Energy.
%%%%%%%%%%%%%%%%%%%%%%%%%%%

%%%%%%%%%%%%%%%%%%%%%%%%%%%%%%%%%%%%%%%%%%%%%%%%%%%%%%%%%%%%%%%%%%%
\section*{Appendix}

We describe the details of our analysis of Yukawa coupling
unification. 
The first step is to compute gauge couplings
and the third generation fermion masses in the $\overline{DR}$ scheme at
the $Z$-boson mass scale $m_Z$, using the full standard model (SM). 

For gauge couplings, we adopt experimental values~\cite{Abbaneo:2001ix}
of the QED fine structure constant $\alpha^{-1}=137.06$, the hadronic
contribution to the QED coupling at $m_Z$ $\Delta \alpha^{(5)}_{\rm had}
(m_Z)=0.02761$, the leptonic effective electroweak mixing angle 
$\sin^2 \theta^{\rm lept}_{\rm eff}=0.23136$, and QCD coupling
$\alpha^{(5)}_s(m_Z)=0.1172$ as input parameters, then we calculate
the $\overline{MS}$ gauge couplings ($g_i^{SM~ \overline{MS}}(m_Z)$)
by using the formula in Refs.~\cite{Fanchiotti:1992tu}. 
We convert them into the $\overline{DR}$ gauge couplings 
($\bar{g}_i^{SM}(m_Z)$)
using the relation between $\overline{MS}$ and $\overline{DR}$
gauge couplings:
\beq
\frac{4\pi}{\bar{g}_i^{SM}(m_Z)^2}
=\frac{4\pi}{g_i^{SM~\overline{MS}}(m_Z)^2}
-\frac{C_i}{12 \pi},
\eeq
where $C_1=0$, $C_2=2$, and $C_3=3$.

For the bottom quark mass,
we adopt the  $\overline{MS}$ bottom quark mass $m_b^{\overline{MS}}$
at $m_b^{\overline{MS}}$ as an input parameter for our
numerical analysis~\cite{Hagiwara:fs}:
\begin{eqnarray}
m_b^{\overline{MS}}(m_b^{\overline{MS}})=4.26\pm0.30~{\rm GeV}.
\label{mb_input}
\end{eqnarray}
Using two-loop (and ${\cal O}(\alpha_s^3)$ for the QCD coupling) 
renormalization group (RG) analysis in the effective QCD and QED
theory, we calculate the $\overline{MS}$ bottom quark mass
at $m_Z$. At $m_Z$ we match the $\overline{MS}$ mass in
the effective QCD and QED theory
into one in the SM ($m_b^{SM~\overline{MS}}(m_Z)$)
including electroweak contribution to the bottom quark.
We then convert it into the $\overline{DR}$ bottom quark mass 
($\bar{m}_b^{SM}(m_Z)$):
\begin{eqnarray}
\bar{m}_b^{SM}(m_Z)&=&m_b^{SM~\overline{MS}}(m_Z)
\left(1-\frac{\alpha_s(m_Z)}{3\pi}-\frac{29\alpha^2_3(m_Z)}{27 \pi^2}
\right),\nonumber \\
 &=& 2.89 \pm0.24~{\rm GeV}
\end{eqnarray}
for $\alpha_s(m_Z)=0.1172$.

For the tau lepton mass, we adopt the pole mass ($m_\tau$) as an input
\cite{Hagiwara:fs}:
\begin{eqnarray}
m_\tau=1776.99^{+0.29}_{-0.26}~{\rm MeV}.
\end{eqnarray}
Then we calculate the $\overline{MS}$ tau mass at $m_\tau$, and
we extrapolate it to one at $m_Z$ using two-loop RG analysis
in the effective QED theory. At $m_Z$, we match it into one in the
SM correcting for the electroweak contributions. Then we get the 
$\overline{DR}$ tau lepton mass at $m_Z$:
\begin{eqnarray}
\bar{m}_\tau^{SM}(m_Z) = 1748.77^{+0.29}_{-0.26}~{\rm MeV}.
\end{eqnarray}

For top quark mass, we use the pole mass ($m_t$) as an input parameter:
\begin{eqnarray}
m_t=174.3\pm 5.1~{\rm GeV}.
\end{eqnarray}
Using a relation between the pole mass and $\overline{DR}$ mass in the
SM at $m_Z$ scale, we compute the $\overline{DR}$ mass at
$m_Z$:
\begin{eqnarray}
\bar{m}_t^{SM}(m_Z)=172.3\pm5.3~{\rm GeV},
\end{eqnarray}
for $\alpha_s(m_Z)=0.1172$.

%%%%%%%%%%%%%%%%%%%%%%%%%%%

In the next step, we compute $\overline{DR}$ couplings in the MSSM at 
$m_Z$. For gauge couplings, we include logarithmic SUSY threshold
corrections\footnote{The SUSY finite corrections will be small if
SUSY particle masses are more than a factor of a few larger than $m_Z$.}.
The relation between the SM and MSSM $\overline{DR}$ gauge
couplings is 
\begin{eqnarray}
\frac{4\pi^2}{\bar{g}_i^{MSSM}(m_Z)^2} =
\frac{4\pi^2}{\bar{g}_i^{SM}(m_Z)^2} +\Delta_i^{SUSY}.
\label{MSSM_gauge_cc}
\end{eqnarray}
Here $\Delta_i^{SUSY}$ is the SUSY logarithmic corrections:
\begin{eqnarray}
\Delta_3^{SUSY} &=& \log{\frac{m_{\tilde{g}}}{m_Z}}
+\frac{N_g}{3}\log{\frac{m_{\tilde{q}}}{m_Z}},
\nonumber \\
\Delta_2^{SUSY} &=& \frac{2}{3}\log{\frac{m_{\tilde{W}}}{m_Z}}
+\frac{N_g}{4}\log{\frac{m_{\tilde{q_L}}}{m_Z}}
+\frac{N_g}{12} \log{\frac{m_{\tilde{l}_L}}{m_Z}}
\nonumber \\
&&
+\frac{1}{3}\log{\frac{m_{\tilde{H}}}{m_Z}}
+\frac{1}{12}\log{\frac{m_H}{m_Z}},\nonumber \\
\Delta_1^{SUSY} &=& \frac{N_g}{60}\log{\frac{m_{\tilde{q}_L}}{m_Z}}
+\frac{2N_g}{15} \log{\frac{m_{\tilde{u}_R}}{m_Z}}
+\frac{N_g}{30}\log{\frac{m_{\tilde{d}_R}}{m_Z}}
\nonumber \\
&&+\frac{N_g}{20}\log{\frac{m_{\tilde{l}_L}}{m_Z}}
+\frac{N_g}{10}\log{\frac{m_{\tilde{e}_R}}{m_Z}}
+\frac{1}{5}\log{\frac{m_{\tilde{H}}}{m_Z}}
+\frac{1}{20}\log{\frac{m_H}{m_Z}}
\label{MSSM_gauge_corr}
\end{eqnarray}

%%%%%%%%%%%%%%%%%%%%%%%%%%%

%\subsection*{$\overline{DR}$ Yukawa couplings at electroweak scale}

In order to calculate the $\overline{DR}$ Yukawa couplings, we need
the $\overline{DR}$ vacuum expectation value (vev) in the MSSM.
In this paper, we define the $\overline{DR}$ vev $(\bar{v}(\mu))$
from the $\overline{DR}$ Z-boson mass:
\begin{eqnarray}
\bar{m}^2_Z(\mu) =\frac{\bar{g}^{'2}(\mu)+\bar{g}_2^2(\mu)}
{4} \bar{v}^2(\mu).
\end{eqnarray}
The pole mass of the $Z$-boson $(m_Z)$ is given by the $\overline{DR}$ mass
and the transverse part of the $Z$-boson self-energy, so we can 
calculate the $\overline{DR}$ mass:
\begin{eqnarray}
\left[p^2-\bar{m}_Z^2(\mu)+\rm{Re}\Pi_Z^{\rm T}(p^2)\right]|
_{p^2=m_Z^2}&=&0, \nonumber \\
\bar{m}_Z^2(\mu)&=&m_Z^2(1+\delta_Z(\mu)),\\
\delta_Z(\mu) &\equiv& \frac{{\rm Re}\Pi_Z^{\rm T}(m_Z^2)}{m_Z^2}.
\end{eqnarray}
In the MSSM, one loop contributions to $\delta_Z$ are 
given by, for example, Eq.~(D.4) in Ref.~\cite{Pierce:1996zz}. 
For example, taking $m_Z=91.1876$ GeV and $m_h=115$ GeV
and assuming all SUSY particle
masses to be $M_{SUSY}=1$ TeV (neglecting all mass mixings),
we get
\begin{eqnarray}
\bar{v}(m_Z)=249.5~{\rm GeV}.
\end{eqnarray}
We note that the SUSY scale $(M_{SUSY})$ dependence of $ \bar{v}$
is very mild because the SUSY corrections to the $\overline{DR}$ $Z$-mass are 
partially canceled by those from the gauge couplings.
For $M_{SUSY}=200~{\rm GeV}-3~{\rm TeV}$ and $m_h=115$ GeV,
\begin{eqnarray}
\bar{v}(m_Z)&=&249.5\pm 0.2~{\rm GeV}
=249.5(1\pm0.08\%)~{\rm GeV}.
\end{eqnarray}
The Higgs mass dependence is also very small:
$\bar{v}(m_Z)=249.5\pm 0.1~{\rm GeV}$ for $m_h=100-135$ GeV.
Therefore, in section 3, we fix $M_{SUSY}=1$ TeV
and $m_h=115$ GeV for the $\bar{v}$.

Then the $\overline{DR}$ Yukawa couplings in the MSSM are
given by
\begin{eqnarray}
\bar{y}_t(m_Z)&=&\frac{\sqrt{2}\bar{m}_t^{MSSM}(m_Z)}
{\bar{v}(m_Z) \sin\beta}=\frac{\sqrt{2}\bar{m}_t^{SM}(m_Z)}
{\bar{v}(m_Z) \sin\beta}(1+\delta_t(m_Z)), \nonumber \\
\bar{y}_b(m_Z)&=&\frac{\sqrt{2}\bar{m}_b^{MSSM}(m_Z)}
{\bar{v}(m_Z) \cos\beta}=\frac{\sqrt{2}\bar{m}_b^{SM}(m_Z)}
{\bar{v}(m_Z) \cos\beta}(1+\delta_b(m_Z)),\nonumber \\
\bar{y}_\tau(m_Z)&=&\frac{\sqrt{2}\bar{m}_\tau^{MSSM}(m_Z)}
{\bar{v}(m_Z) \cos\beta}=\frac{\sqrt{2}\bar{m}_\tau^{SM}(m_Z)}
{\bar{v}(m_Z) \cos\beta}(1+\delta_\tau(m_Z)).
\label{DR-yukawa}
\end{eqnarray}
where $\delta_f(m_Z)$ are the weak-scale corrections due to
SUSY particle loops.
A relation between the fermion pole ($m_f$) and $\overline{DR}$
($\bar{m}_f(\mu)$) masses is defined by
\begin{eqnarray}
\left.[\slash\!\!\!{p} 
- \bar{m}_{f}(\mu) +\Sigma (\/ p)]
\right|_{\slash\!\!\!p = m_f}&=&0,
\nonumber \\
\bar{m}_t(\mu)&=&m_f \left\{1+ \frac{\Sigma_f(m_f)}{m_f}\right\}.
\end{eqnarray}
where $\Sigma_f$ is the self-energy of the fermion propagator.
Therefore the SUSY contributions $\delta_f(m_Z)$ ($f=t,b$ and $\tau$) 
are defined
by
\beq
\delta_f(m_Z)=\frac{\Sigma(m_f)^{MSSM}-\Sigma(m_f)^{SM}}{m_f}.
\label{yukawa_correction}
\eeq
The one loop SM and SUSY contributions to $\delta_{t,b,\tau}$ can be
found, for example,  in Eq. (D.18) in Ref.~\cite{Pierce:1996zz}.

After we get all $\overline{DR}$ gauge and Yukawa couplings at $m_Z$ in
the MSSM, we numerically solve two loop RG equations for the full MSSM 
from $m_Z$ scale to the GUT scale in order
to analyze the couplings at the GUT scale, and test for unification.

%%%%%%%%%%%%%%%%%%%%%%%%%%%%%%%%%%%%%%%%%%%%%%%%%%%%%%%%%%%%%%%%%%%

\end{document}